\documentclass[10pt,a4paper,twocolumn,english,aps,prl,twocolumn,showpacs,floatfix,superscriptaddress,10pt]{revtex4-1}
\usepackage[T1]{fontenc}
\usepackage[latin9]{inputenc}
\usepackage{babel}
\usepackage{bm}
\usepackage{array}
\usepackage{float}
\usepackage{multirow}
\usepackage{amssymb}
\usepackage{graphicx}
\usepackage[normalem]{ulem}

\usepackage{xcolor}
\definecolor{grey}{rgb}{.6,.6,.6}

\PassOptionsToPackage{normalem}{ulem}
\usepackage[unicode=true,
 bookmarks=true,bookmarksnumbered=true,bookmarksopen=true,bookmarksopenlevel=2,
 breaklinks=false,pdfborder={0 0 0},backref=false,colorlinks=true]
 {hyperref}
\hypersetup{pdfauthor={Maxen Cosset-Chéneau },pdfsubject={Spintronics}}
\usepackage{breakurl}



\begin{document}

\title{Measurement of the Spin Absorption Anisotropy in Lateral Spin Valves}

\author{M. Cosset-Ch\'{e}neau}
\affiliation{Universit\'{e} Grenoble Alpes, CEA, CNRS, INP-G, Spintec, F-38054 Grenoble, France}

\author{L. Vila}
\affiliation{Universit\'{e} Grenoble Alpes, CEA, CNRS, INP-G, Spintec, F-38054 Grenoble, France}

\author{G. Zahnd}
\affiliation{Universit\'{e} Grenoble Alpes, CEA, CNRS, INP-G, Spintec, F-38054 Grenoble, France}

\author{D. Gusakova}
\affiliation{Universit\'{e} Grenoble Alpes, CEA, CNRS, INP-G, Spintec, F-38054 Grenoble, France}

\author{V.T. Pham}
\affiliation{Universit\'{e} Grenoble Alpes, CEA, CNRS, INP-G, Spintec, F-38054 Grenoble, France}

\author{C. Gr\`ezes}
\affiliation{Universit\'{e} Grenoble Alpes, CEA, CNRS, INP-G, Spintec, F-38054 Grenoble, France}

\author{X. Waintal}
\affiliation{Universit\'{e} Grenoble Alpes, CEA, Pheliqs, F-38054 Grenoble, France}

\author{A. Marty }
\affiliation{Universit\'{e} Grenoble Alpes, CEA, CNRS, INP-G, Spintec, F-38054 Grenoble, France}

\author{H. Jaffr\`es }
\affiliation{Unit\'{e} Mixte de Physique CNRS/Thales, university Paris-Sud and Universit\'{e} Paris-Saclay, 91767 Palaiseau, France}

\author{J.-P. Attan\'{e} }
\affiliation{Universit\'{e} Grenoble Alpes, CEA, CNRS, INP-G, Spintec, F-38054 Grenoble, France}


\date{\today}
\selectlanguage{english}%
\begin{abstract}
The spin absorption process in a ferromagnetic material depends on the spin orientation relatively to the magnetization. Using a ferromagnet to absorb the pure spin current created within a lateral spin-valve, we evidence and quantify a sizeable orientation dependence of the spin absorption in Co, CoFe and NiFe. These experiments allow determining the spin-mixing conductance, an elusive but fundamental parameter of the spin-dependent transport. We show that the obtained values cannot be understood within a model considering only the Larmor, transverse decoherence and spin diffusion lengths, and rather suggest that the spin-mixing conductance is actually limited by the Sharvin conductance.
\end{abstract}

\maketitle

The absorption of spin-currents at ferromagnetic/non magnetic interfaces is a fundamental ingredient of many spintronic phenomena such as the spin transfer torque \cite{Berger1995,Slonczewski2002,Ralph2008} and the spin-orbit torques (SOT) \cite{Manchon2019, Miron2010}. In both cases, irrespectively of their initial orientations, the injected spins eventually align with the local magnetization, and as the corresponding angular momentum is transferred to the magnetization, a torque is exerted. Several magnetoresistance effects, from giant magnetoresistance to spin-Hall magnetoresistances (SMR)\cite{Chen2013}, also involve the spin dependent absorption/reflection at interfaces with ferromagnets. 

The relaxation of incoming spins on the local magnetization occurs \textit{via} different relaxation processes depending on the orientation
of the spin with respect to the local magnetization. On one hand, the longitudinal spin component is absorbed over the spin diffusion length $l_{sf}$. As $l_{sf}$ is larger than the mean free path $l_*$ of the material, this phenomena is well described by the Valet-Fert diffusive model \cite{VF1993}. The absorption of the longitudinal spin current is  proportional to the spin-flip rate, and thus inversely proportional to the material spin resistance $R_s$ \cite{Laczkowski2015}. On the other hand, the transverse component of the spin, relaxes on a very short (ballistic) length scale associated with the Larmor precession of the conduction electrons spins around the strong \textit{s-d} exchange field of the ferromagnet \cite{Petitjean2012} or around the local spin-orbit field. This leads to decoherence, and possibly to the appearance of spin transfer torques \cite{Stiles2002}. Another source of  transverse spin relaxation is the conduction band mismatch at the interface between the two materials, which leads to spin-dependent transmission and reflection processes~\cite{Stiles2002, Berger1995}. Both these ballistic effects likely play a role in the absorption of the transverse component \cite{Petitjean2012}. As the characteristic lengths involved in the transverse case are expected to be smaller (typically 1 to 2~nm \cite{Taniguchi2008,Brataas2006}) than in the longitudinal case \cite{Zahnd2018} (typically 8~nm in Co and 5~nm in NiFe), one may anticipate a much larger absorption efficiency for the transverse spin component.

As transverse spins get absorbed on a very short length scale, the description of the absorption process only requires the addition of a single parameter with respect to the Valet-Fert theory: the interfacial spin-mixing conductance \cite{Brataas2006}. However, and despite attempts using the Hanle effect and Spin Hall Magnetoresistance \cite{Idzuchi2014, Weiler2013}, this parameter remains very difficult to measure experimentally \cite{Zhu2019}.

\begin{figure}[h]
    \centering
    \includegraphics[scale=0.55]{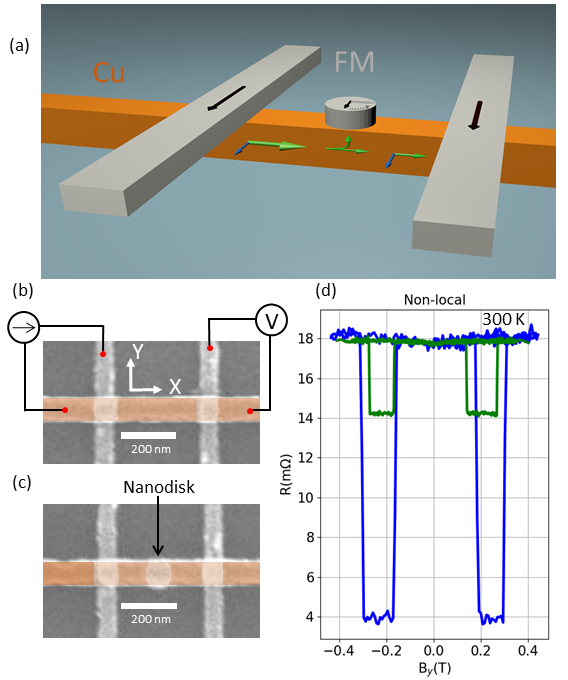}
    \caption{(a) Magnetization (black and gray arrows) and spin current (green arrows) in the LSV, with the Cu in orange and the ferromagnetic material in grey. 
     The green arrows represent the spin current direction and the blue arrows its spin polarization. The ferromagnetic nanodisk absorbs part of the pure spin current created by the left electrode. The two perpendicular arrows on this absorber indicate two possible magnetization orientations, collinear or transverse to the injected spins. The spin signal, measured on the right side of the device, depends on the relative orientation of this magnetization with the injected spins. (b) Colored Scanning Electron Microscopy (SEM) image of a reference LSV, i.e., without absorber. The orange part corresponds to the Cu channel, and the grey to the ferromagnetic electrodes. (c) SEM image of a lateral spin valve with a ferromagnetic absorber. The ferromagnetic elements are all 30~nm thick. The ferromagnetic electrodes are 50~nm wide and the absorber has a diameter of 80~nm. The center-to-center distance between the source and drain is 300 nm. The non-magnetic Cu channel is 80~nm thick and 80~nm wide. (d) Comparison of the non-local measurement for a reference LSV and for a LSV with absorber, when applying a magnetic field along the electrodes easy magnetization axis (Y-axis).}
    \label{fig:fig1}
\end{figure}

In this letter, we present the experimental evidence of a strong spin absorption anisotropy in 3d ferromagnets, using a lateral spin valve (LSV) with a nanodisk-shaped magnetic absorber (cf. \ref{fig:fig1}a). Using the non-local measurement configuration shown in \ref{fig:fig1}b, a pure spin current is created within the Cu channel of a LSV, with a spin orientation along Y. A ferromagnetic nanodisk placed along the channel acts as an absorber of the spin current. The output spin signal of the LSV then varies with the orientation of the absorber magnetization, which is controlled by an external magnetic field. The anisotropy of the spin absorption is found to be significant, with spin signal variations of around 40\%. We use a dedicated spin transport model to extract from these data the spin mixing conductance of the interfaces between Cu and Co, CoFe or NiFe. 
Finally, the comparison of our experimental results with previous ab-initio calculations indicates that the relaxation of the transverse component cannot be understood without considering the existence of an upper limit to the spin-mixing conductance given by the Sharvin conductance.

In order to obtain large spin signals at room temperature, the ferromagnetic source and detector of the LSVs are made of CoFe \cite{Zahnd2016}. The non-magnetic channel is made of Cu, in order to take benefit of its long spin-diffusion length, and of rather small resistances of the interfaces with ferromagnetic parts \cite{Pratt2009}. The absorber is formed by a nanodisk made of Co, CoFe, or NiFe. In contact with the channel, it acts as a spin sink, \textit{i.e}., a pure spin current flows towards the absorber to relax the spin accumulation. LSVs without absorber, elaborated in the same batch, are used as non-local MR signal references (fig. \ref{fig:fig1}b). 

The devices have been patterned on PMMA by e-beam lithography on a $\mathrm{SiO}_2$ substrate, followed by physical vapor deposition and lift-off. The nanodisk has been deposited in a first step. The ferromagnetic electrodes and the non-magnetic channel have been deposited during the second and third steps, respectively. Argon ion beam milling is used before Cu deposition, in order to obtain clean interfaces. The transparency of the interfaces in our devices~\cite{Pham2016,Zahnd2018n2} is large, giving rise to a small interface resistance of the order of 1 f$\Omega.m^2$, consistent to those obtained in Ref.~\cite{Stiles2000}, and corresponding to the smallest values that can be achieved in disordered interfaces \cite{Brataas2006}. We also measured in theses devices large effective 
polarizations~\cite{Zahnd2016,Laczkowski2019} and record giant magnetoresistances (more than 10\%~\cite{Zahnd2017}) in lateral devices, which indicates that the spin memory loss at the interface can be neglected. 
The resistances, spin diffusion lengths and spin polarizations of the different materials are given in Supplemental Materials.

Magnetoresistance measurements have been performed using lock-in techniques (I=$300 \mu$~A; f=$330$~Hz). All measurements have been performed at 300K. The probing configuration, shown in fig.~1b, is that of a non-local measurement~\cite{Takahashi2003}. In a reference LSV, without absorber, a current flowing through the left-side ferromagnetic/non-ferromagnetic interface (the injector) generates a spin accumulation, and thus creates a pure spin current along the non-magnetic channel, with a spin polarization vector collinear to the injector magnetization. At the detecting electrode, on the right-hand side, the remaining spin-accumulation yields a voltage drop at the CoFe/Cu interface~\cite{Takahashi2003}. Such a non-local probe technique allows avoiding spurious effects, as the spin-current along the channel is pure, \textit{i.e.}, free of any charge flow ~\cite{Takahashi2003}. When adding an absorber to the LSV (fig.~\ref{fig:fig1}c), the spin current is partly absorbed by the nanodisk. 
The parallel and antiparallel magnetization states between the two electrodes correspond to a high and low voltage drop at the interface, respectively~\cite{Otani2011}. 
The spin signal amplitude, $\Delta R$ (in Ohms), corresponds to the difference of potential between these two states, divided by the injected charge current. As seen in fig.~\ref{fig:fig1}a, when an absorber is inserted it acts as a sink for the spin accumulation, diverting part of the spin current. As expected, this leads for all materials to an important decrease (70-80\%) of the spin signal amplitude (~\ref{fig:fig1}d) with respect to the reference LSV\cite{Zahnd2018}.

\begin{figure}[h]
    \centering
    \includegraphics[scale=0.5]{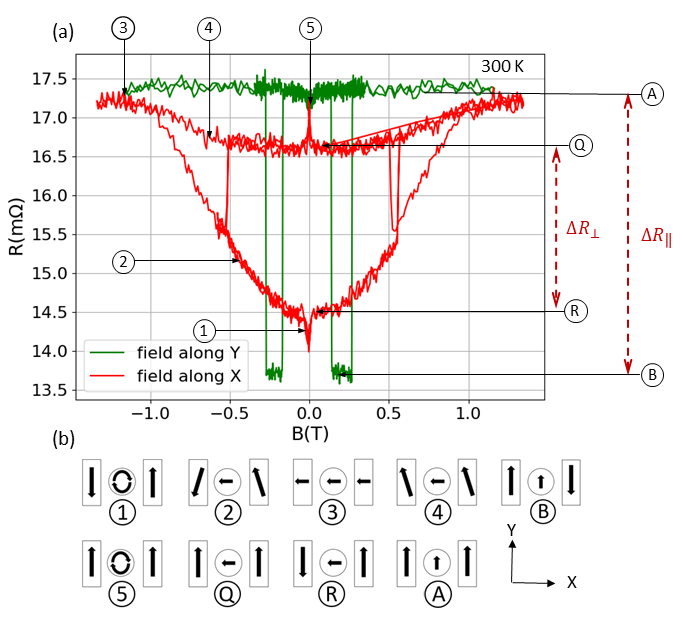}
    \caption{(a) Non-local measurement of the LSV with absorber when applying a magnetic field along the Y (green) and X-axis (red). (b) Magnetic configurations of the injector, absorber and detector, for magnetic fields applied along the X-axis.}
    \label{fig:fig2}
\end{figure}

Measurements acquired for magnetic fields applied along the X or Y direction are shown on Fig.~\ref{fig:fig2}a, with the corresponding states of magnetization of the LSV reported on Fig.~\ref{fig:fig2}b.
At zero field, the magnetization of the absorber is in a vortex state (cf. state 1 of fig.~\ref{fig:fig2}b, see also micromagnetic calculations in Supplemental Material). A relatively small in-plane magnetic field (typically 0.1T) saturates the nanodisk magnetization, while the magnetization of the electrodes is left essentially unchanged because of the shape anisotropy (cf. states Q,R,A,B of fig.~\ref{fig:fig2}b). 

For large fields along X, the magnetization of the electrodes rotates towards X, with a saturation at around 1T (state 3). For large fields along Y, one observe the sudden magnetization switching of the electrodes at 0.2 and 0.3T, respectively.

The transverse magnetic configuration, in which the incoming spins are transverse to the magnetization of the absorber, matches with the states Q and R of Fig.~\ref{fig:fig2}b. The spin signal difference between these two states corresponds to the so-called \textit{transverse} spin signal amplitude $\Delta R_\perp=R_Q-R_R$. As displayed on Fig.~\ref{fig:fig2}a, this transverse spin signal amplitude is notably smaller than the \textit{collinear} spin signal amplitude $\Delta R_{//}=R_A-R_B$. 

\begin{figure}[h]
    \centering
    \includegraphics[scale=0.6]{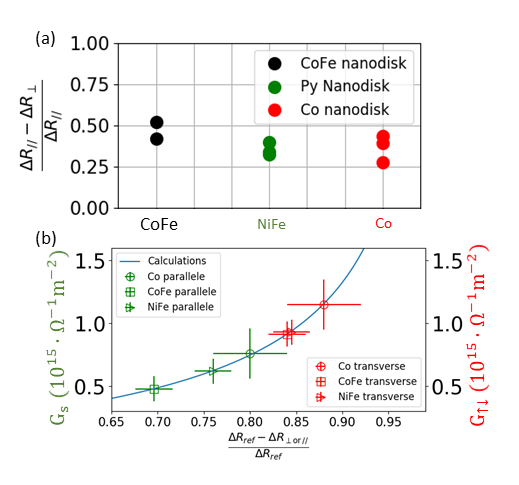}
    \caption{(a) Relative decrease of the non-local spin signal amplitude from the parallel to the transverse configuration. Each symbol corresponds to a single device. (b) Relative variation of the spin signal amplitude when inserting an absorber, as a function of the interface conductance $G=G_s$ (in the collinear case, in green) or $G_{\uparrow \downarrow}$ (in the transverse case, in red). Different symbols correspond to different materials. The blue curve has been obtained with the model proposed in (cf. Supplemental Material), and is valid for both the collinear and transverse case. The absorption values are experimental. The error bars reflects the experimental dispersion of the absorption from device to device}
    \label{fig:fig3}
\end{figure}

Changing the magnetization orientation of the absorber by application of a small magnetic field leads to a large modulation of the spin signal, by about $(\Delta R_{//}-\Delta R_{\perp})/\Delta R_{//} \sim 40 \%$. 
As shown in Fig~\ref{fig:fig3}a, the spin signal variations between the parallel and transverse configurations are of the same order of magnitude for the three different materials. Note that this has been observed for each material in several sets 
of devices, with a good reproducibility. Moreover, using micromagnetic simulations, we have checked that the decrease of the spin signal in the transverse configuration does not originate from the Hanle effect around the external or the demagnetizing magnetic field (see Supplemental Material).

This variation of spin signal between the transverse and collinear configuration is a clear manifestation of the spin absorption anisotropy in ferromagnets. For all the studied ferromagnetic materials we systematically find the absorption to be more efficient in the transverse configuration.

At the interface between the channel and the absorber, the spin accumulation can be represented by a vector $\Delta \boldsymbol{\mu}$. It leads to the occurrence of a pure spin current diffusing along $z$, $\boldsymbol{j}_z^s$, whose three components give the spin polarization direction and amplitude. The relationship between the spin accumulation and the spin current is given by \cite{Chen2013}:
\begin{equation}
    \boldsymbol{j}_z^s=G_s (\boldsymbol{m}\cdot \Delta \boldsymbol{\mu})\boldsymbol{m}
    +G_{\uparrow\downarrow} \boldsymbol{m}\times (\Delta \boldsymbol{\mu}\times \boldsymbol{m}) 
\end{equation}
where $\boldsymbol{m}$ is the unit magnetization vector of the absorber. The absorption of the collinear component of the spin current is described by the spin conductance $G_s$, and $G_{\uparrow\downarrow}$ stands for the spin mixing conductance relevant for the two non-collinear spin components. Here we neglect its imaginary part, typically one order of magnitude smaller than its real part for metallic interfaces \cite{Brataas2006}. These quantities may describe both ballistic and diffusive aspects, and control the dependence of the spin absorption upon the spin direction. Although difficult to measure with a large precision, $G_s$ and $G_{\uparrow\downarrow}$ are fundamental interface parameters, and key values to understand interface-related magnetoresistances, STT, and SOT \cite{Stiles2002}. 

In the following, we propose an analysis of our experimental results allowing to extract $G_s$ and $G_{\uparrow\downarrow}$. To these ends, We combined an advanced analytical approach with discrete numerical calculations to compute the spin signal as a function of $G_s$ and $G_{\uparrow\downarrow}$ (cf. Supplemental Material). The method consists in solving the equations proposed by Petitjean et al. \cite{Petitjean2012}, adapted to the case of large ferromagnetic thicknesses, in order to obtain the spin current within the device geometry. Apart from $G_s$ and $G_{\uparrow\downarrow}$, considered as free parameters, the different material properties used for these calculations are extracted independently from lateral spin-valves measurements \cite{Zahnd2018}. 

In fig. ~\ref{fig:fig3}b we use our theoretical modeling (blue curve) to extract the values of $G_s$ (colinear case) or $G_{\uparrow \downarrow}$ (perpendicular case) from our experimental measurement of $\Delta R$.  The results are summarized in Table~\ref{table:1}. The variations of $(\Delta R_{ref}-\Delta R_{//})/\Delta R_{ref}$ on $G_s$ and of $(\Delta R_{ref}-\Delta R_{\perp})/\Delta R_{ref}$ on $G_{\uparrow\downarrow}$ predicted by our model are found to be identical, $\Delta R_{ref}$ being the spin signal amplitude of a LSV without absorber. These variations are thus represented by the single blue curve. This particular feature may be understood by reminding that the imaginary part of $G_{\uparrow\downarrow}$ has been neglected, and that the absorber thickness is larger than the relaxation lengths (cf. Supplemental Material). The position of the symbols along the Abscissa corresponds to the measured absorption efficiency. For the collinear configuration ($(\Delta R_{ref}-\Delta R_{//})/\Delta R_{ref}$), from the blue curve, one then can have access to the corresponding value of $G_s$ in our system. Similarly, $G_{\uparrow\downarrow}$ can be deduced from the absorption $(\Delta R_{ref}-\Delta R_{\perp})/\Delta R_{ref}$ acquired in the transverse case. The absorption being enhanced in the transverse configuration, $G_{\uparrow\downarrow}$ is larger than $G_s$. The ensemble of values extracted is gathered in table~\ref{table:1}. In the collinear case, we will conclude that $G_s$ is similar to that measured previously in lateral spin-valves \cite{Zahnd2018}. In the transverse configuration, the values of $G_{\uparrow \downarrow}$ extracted are of the same range of magnitude than the few existing data, either given by SP-FMR~\cite{Ghosh2012} or by Hanle effect experiments~\cite{Idzuchi2014}.

Let us now compare the experimental conductances to the theoretical predictions, first considering a simplified picture based only on the relaxation lengths such as proposed in Ref.~\cite{Nonoguchi2012}. The main hypothesis are the following: i) the interface is purely transparent, ii) the relaxation of the collinear spin component is controlled by the rate of spin-flip scaling with the inverse of the absorber spin diffusion length \cite{VF1993, Laczkowski2015}, and iii) the relaxation of the transverse spin component is mainly controlled by the correlated Larmor length, $l_L$, and the transverse spin relaxation length, $l_\perp$~\cite{Petitjean2012}. In these hypothesis, $G_s$ depends on $\lambda_F$, the spin diffusion length of the absorber. Here, the thickness of the absorber is much larger than the typical spin diffusion length $\lambda_F$ of 3\text{d} ferromagnets ~\cite{Zahnd2018}. Consequently, in the collinear case the efficiency of the absorption scales with an effective spin conductance, $G_s=1/R_s$, inverse of the spin resistance $R_s=\rho_F \lambda_F/(1-\beta_F^2)$, where $\beta_F$ is the absorber polarization, and $\rho_F$ its resistivity. 

Using data from previous experiments ~\cite{Zahnd2018}, $G_s$ is deduced to be close to $0.61^{+0.26}_{-0.20}\cdot 10^{15}$, $0.62^{+0.05}_{-0.08}\cdot 10^{15}$ and $0.5^{+0.25}_{-0.16}\cdot 10^{15}\, \Omega^{-1}$~m$^{-2}$ for NiFe, CoFe and Co respectively. As one may note, these values are consistent with those obtained here using the blue curve of Fig.~\ref{fig:fig3}b. On the other hand, in the transverse case, within our simplified picture of relaxation length dependence, the spin mixing conductance denoted here $G_{\uparrow \downarrow}^{L}$ (where L stands for 'Larmor') may be estimated at $G_{\uparrow \downarrow}^{L} \approx 1/[\left(\min(l_L, l_\perp) \rho_F\right]$, since $\lambda_F\gg l_L,l_\perp$. According to ab-initio calculations ~\cite{Petitjean2012}, this leads to $G_{\uparrow \downarrow}^{L}$ values ranging from $3\cdot10^{15}\, \Omega^{-1}$~m$^{-2}$ for NiFe to $13\cdot 10^{15} \, \Omega^{-1}$~m$^{-2}$ for Co.
In the absence of computed values for $l_L$ and $l_\perp$ in CoFe, no prediction can be made for the value of $G_{\uparrow\downarrow}$. However, by considering the values given for Co and Fe in Ref.~\cite{Petitjean2012} it seems safe to assume that $G_{\uparrow \downarrow}^{L}$ is larger than $3\cdot10^{15} \, \Omega^{-1}$~m$^{-2}$.

This simplified approach, in which the sole transport lengths determine the spin absorption, predicts values of $G_s$ (\textit{c.f.} table~\ref{table:1}) close to our measurements. However, those appear much larger than the experimental ones which goes in favor of additional dependence than the spin diffusion, Larmor precession and transverse decoherence lengths. We suggest in the following that, even at room temperature, one needs to consider a ballistic contribution. As the density of conduction channels in Cu is finite, the transverse spin absorption is somehow limited. The physical quantity associated to this quantum limitation of the interface conductivity is the Sharvin conductance \cite{Bauer2002}, and although a key parameter of transport, it seems rarely taken into account in pure spintronics experiments \cite{Zhu2019}. 

Up to a small correction that characterises the spin dependence of the reflection at a given interface, the spin-mixing conductance is expected to be rather close to the Sharvin conductance \cite{Brataas2006}. Here, our experimental values of $G_{\uparrow \downarrow}$ are indeed comparable to the Sharvin conductance $G_{sh}=1.2\cdot 10^{15}~\Omega^{-1}$~m$^{-2}$ of 3\textit{d}/Cu interfaces obtained by ab-initio methods~\cite{Brataas2006}. 

\begin{table}[h!]
\centering
\begin{tabular}{|c| c c c|} 
 \hline
     & Cu/CoFe & Cu/Co & Cu/NiFe \\ [0.5ex] 
 \hline
 \hline
 $G_s \, (10^{15} \Omega^{-1}m^{-2})$ & $0.48\pm0.2$ & $0.76\pm0.1$ & $0.62\pm0.1$ \\ 
 $G_{\uparrow\downarrow}  \, (10^{15} \Omega^{-1}m^{-2})$ & $0.91\pm0.1$ & $1.1\pm0.2$ & $0.93\pm0.1$ \\ [1ex] 
 \hline
\end{tabular}
\caption{Extracted spin conductances $G_s$ and effective spin mixing conductances $G_{\uparrow\downarrow}$ of the Cu/F interface between the channel and the absorbing nanodisk, at room temperature. These values have been extracted from the experimental absorption efficiencies and by using the blue curve of Fig.~\ref{fig:fig3}b.}
\label{table:1}
\end{table}

We can thus explain our results in a simplified manner by taking into account in series both the Sharvin conductance $G_{sh}$ and the real part of the spin mixing conductance 
$G_{\uparrow \downarrow}^{L}$ 
within the \textit{relaxation lengths picture}, so that the overall spin mixing 
conductance writes \textit{in fine} $1/G_{\uparrow \downarrow}=1/G_{sh}+1/G_{\uparrow \downarrow}^{L}$.  According to previous ab-initio calculations \cite{Brataas2006}, the Sharvin conductance is typically 3-10 times smaller than the estimations of $G_{\uparrow \downarrow}^{ L}$, \textit{i.e.}, $G_{\uparrow \downarrow}^{L}\gg G_{sh}$. The spin absorption in the transverse case is thus actually limited by the Sharvin conductance, and the experimental values of $G_{\uparrow \downarrow}$ remains close to the values of $G_{sh}$ in Cu.

Note that, as proposed by K.~S. Das et al. \cite{Das2020}, our device can be also used as a spin transistor, whereby the absorber constitutes a magnetic gate allowing a significant modulation of the spin signal. The 40\% variations observed here reveal indeed relatively large with respect to the state of the art \cite{Cornelissen2018, Dejene2015}.
This contrast could even be enhanced by increasing the difference between $G_{\uparrow \downarrow}$ and $G_s$. This may be obtained by minimizing the spin absorption in the collinear case, \textit{e.g.} by using a thin ferromagnet with a long diffusion length. Although this might seem at first counter-intuitive, one has to remind that the idea is not, to maximize the absorption~\cite{Sagasta2017,Isasa2015,Kimura2005,Nonoguchi2012}, but rather to enhance the contrast between the collinear and transverse spin signals.


To conclude, we used LSVs possessing a ferromagnetic absorber to study the absorption of pure spin currents from a Cu channel into Co, CoFe and NiFe, for both the collinear and transverse configurations. The overall absorption is found to be large. Nonetheless, in the transverse geometry a significant reduction of the spin signal is obtained with respect to the parallel case. Using analytical modelling, we were able to extract the spin mixing conductance at the interface between copper and NiFe, CoFe or Co. These values are too small to be understood by considering the sole role of the relaxation lengths (Larmor, transverse decoherence and spin diffusion lengths). Comparisons with ab-initio calculations rather suggest that the absorption of the transverse spin is actually limited by the Sharvin conductance. This upper bound of the conductance might thus play an important role in spintronics, as most magnetoresistances and spin transfer torques involve spin currents crossing non-magnetic/ferromagnetic interfaces.

\begin{acknowledgments}
We acknowledge the financial support by ANR French National Research Agency OISO (ANR-17-CE24-0026). 
Interesting discussions with Albert Fert are gratefully acknowledged
\end{acknowledgments}


\begin{thebibliography}{10}

\bibitem{Berger1995}
L. Berger, 
\newblock "Precession of Conduction-Electron Spins Near an Interface Between Normal and Magnetic Metals"
\newblock {\em IEEE Trans. Magn.  } 31, 3871 (1995)

\bibitem{Slonczewski2002}
J.C. Slonczewski, 
\newblock "Currents and torques in metallic magnetic multilayers"
\newblock {\em J. Magn. Magn. Mater.} 247, 324-338 (2002)

\bibitem{Ralph2008}
D.C. Ralph, M.D. Stiles, 
\newblock "Spin transfer torque"
\newblock {\em J. Magn. Magn. Mater.}  320 (2008)

\bibitem{Manchon2019}
A. Manchon, J. Zelezny, I.M. Miron, T. Jungwirth, J. Sinova, A. Thiaville, K. Garello, and P. Gambardella
\newblock "Current-induced spin-orbit torques in ferromagnetic and antiferromagnetic systems",
\newblock {\em Rev. Mod. Phys. } 91, 035004 (2019)

\bibitem{Miron2010}
Ioan Mihai Miron, Gilles Gaudin, Stephane Auffret, Bernard Rodmacq, Alain Schuhl, Stefania Pizzini, Jan Vogel and Pietro Gambardella
\newblock "Current-driven spin torque induced by the Rashba effect in a ferromagnetic metal layer",
\newblock {\em Nature Mater. } 9, 230 (2010)

\bibitem{Chen2013}
Y.-T. Chen, S. Takahashi, H. Nakayama, M. Althammer, S. T. B. Goennenwein, E. Saitoh, and G. E. W. Bauer,  
\newblock "Theory of spin Hall magnetoresistance",
\newblock {\em Phys. Rev. B } 87, 144411 (2013)


\bibitem{VF1993}
Valet, T., and A. Fert, 
\newblock "Theory of the perpendicular magnetoresistance in magnetic multilayers"
\newblock {\em Phys. Rev. B } 48, 7099 (1993)

\bibitem{Laczkowski2015}
P. Laczkowski et al,  
\newblock "Evaluation of spin diffusion length of AuW alloys using spin absorption experiments in the limit of large spin-orbit interactions",
\newblock {\em Phys. Rev. B } 92, 214405 (2015)

\bibitem{Petitjean2012}
C. Petitjean, D. Luc, X. Waintal, 
\newblock "Unified Drift-diffusion Theory for Transvserse Spin currents in Spin Valves, Domain Walls, and Other Textured Magnets"
\newblock {\em Phys. Rev. Lett.} 109, 117204 (2012)

\bibitem{Stiles2002}
M.D. Stiles and A. Zangwill, 
\newblock "Anatomy of spin transfer torque"
\newblock {\em Phys. Rev. B } 66, 014407 (2002)

\bibitem{Taniguchi2008}
T. Taniguchi, S. Yakata, H. Imamura, Y. Ando, 
\newblock "Determination of penetration depth of transverse spin current in ferromagnetic metals by spin pumping"
\newblock {\em Appl. phys. express} 1, 031302 (2008)


\bibitem{Brataas2006}
A. Brataas, G.E. Bauer, P.J. Kelly, 
\newblock "Non-collinear magnetoelectronics"
\newblock {\em Rev. Mod. Phys} 427, 157 (2006)

\bibitem{Idzuchi2014}
Idzuchi, H., et al,  
\newblock "Effect of anisotropic spin absorption on the Hanle effect in lateral spin valves",
\newblock {\em Phys. Rev. B } 89, 081308 (2014)

\bibitem{Weiler2013}
M. Weiler, M. Althammer, M. Schreier, J. Lotze, M. Pernpeintner, S. Meyer, H. Huebl, R. Gross, A. Kamra, J. Xiao, Y.-T. Chen, H. Jiao, G. E. W. Bauer, and S. T. B. Goennenwein,
\newblock "Experimental Test of the Spin Mixing Interface Conductivity Concept",
\newblock {\em  Appl. Phys. Lett. } 111, 176601 (2013)

\bibitem{Zhu2019}
L. Zhu, D. C. Ralph, and R. A. Buhrman,
\newblock "Effective Spin-Mixing Conductance of Heavy-Metal-Ferromagnet 
Interfaces",
\newblock {\em Phys. Rev. lett.} 123, 057203 (2019)

\bibitem{Zahnd2018}
G. Zahnd, L. Vila, V. T. Pham, M. Cosset-Cheneau, W. Lim, A. Brenac, P. Laczkowski, A. Marty, and J. P. Attane, 
\newblock "Spin diffusion length and polarization of ferromagnetic metals measured by the spin-absorption technique in lateral spin valves"
\newblock {\em Phys. Rev. B } 98, 175514 (2018)

\bibitem{Das2020}
K. S. Das, F. Feringa, M. Middelkamp, B. J. van Wees, and I. J. Vera-Marun, 
\newblock "Modulation of magnon spin transport in a magnetic gate transistor"
\newblock {\em Phys. Rev. B } 101, 054436 (2020)

\bibitem{Cornelissen2018}
L. J. Cornelissen, J. Liu, B. J. van Wees, and R. A. Duine 
\newblock "Spin-Current-Controlled Modulation of the Magnon Spin Conductance in a Three-Terminal Magnon Transistor"
\newblock {\em Phys. Rev. Lett. } 120, 097702 (2018)

\bibitem{Dejene2015}
F. K. Dejene, N. Vlietstra, D. Luc, X. Waintal, J. Ben Youssef, and B. J. van Wees, 
\newblock "Control of spin current by a magnetic YIG substrate in NiFe/Al nonlocal spin valves"
\newblock {\em Phys. Rev. B } 91, 100404(R) (2015)

\bibitem{Zahnd2016}
G. Zahnd, L. Vila, T.V. Pham, A. Marty, P. Laczkowski, W. Savero Torres, C. Vergnaud, M. Jamet and J.P. Attane, 
\newblock "Comparison of the use of NiFe and CoFe as electrodes for metallic lateral spin valves",
\newblock {\em Nanotechnology } 27, 035201 (2016)

\bibitem{Pratt2009}
W.P. Pratt Jr., J. Bass,  
\newblock "Perpendicular-current studies of electron transport across metal/metal interfacess",
\newblock {\em Appl. Surf. Sci. } 256, 399 (2009)

\bibitem{Pham2016}
V.T. Pham, G. Zahnd, A. Marty, W. Savero Torres, M. Jamet, P. Noel, L. Vila and J.P. Attane,  
\newblock "Electrical detection of magnetic domain walls by inverse and direct spin Hall effect",
\newblock {\em Appl. Phys. Lett.} 109, 192401 (2016)

\bibitem{Zahnd2018n2}
G. Zahnd, L. Vila, V.T. Pham, F. Rortais, M. Cosset-Cheneau, C. Vergnaud, M. Jamet, P. Noel, T. Gushi, A. Brenac, A. Marty, and J.P. Attane,  
\newblock "Observation of the Hanle effect in giant magnetoresistance measurements",
\newblock {\em Appl. Phys. Lett.} 112, 232405 (2018)

\bibitem{Stiles2000}
Stiles, Mark D., and David R. Penn,  
\newblock "Calculation of spin-dependent interface resistance",
\newblock {\em Phys. Rev. B } 61, 3200 (2000)

\bibitem{Laczkowski2019}
P. Laczkowski, M. Cosset-Cheneau, W. Savero-Torres, V. T. Pham, G. Zahnd, H. Jaffres, N. Reyren, J.-C. Rojas-Sanchez, A. Marty, L. Vila, J.-M. George, and J.-P. Attane,  
\newblock "Spin-dependent transport characterization in metallinc lateral spin valves using one-dimensional and three-dimensional modeling",
\newblock {\em Phys. Rev. B }  99, 134436 (2019)

\bibitem{Zahnd2017}
G. Zahnd, L. Vila, V.T. Pham, A. Marty, C. Beigne, C. Vergnaud and J.P. Attane,  
\newblock "Giant magnetoresistance in lateral metallic nanostructures for spintronic applications",
\newblock {\em Sci. Rep.} 7, 9553 (2017)

\bibitem{Takahashi2003}
S. Takahashi and S. Maekawa,  
\newblock "Spin injection and detection in magnetic nanostructures",
\newblock {\em Phys. Rev. B} 67, 052409 (2003)

\bibitem{Otani2011}
Y. Otani and T. Kimura,
\newblock "Manipulation of spin currents in metallic systems"
\newblock {\em Philos. Trans. R. Soc. London A } 369, 3136 (2011) 

\bibitem{Jaffres2010}
 H. Jaffres, J.-M. George, and A. Fert, 
\newblock "Spin transport in multiterminal devices: Large spin signals in
devices with confined geometry",
\newblock {\em Phys. Rev. B } 82, 140408 (2010)

\bibitem{Nonoguchi2012}
S. Nonoguchi, T. Nomura and T. Kimura,  
\newblock "Longitudinal and transverse spin current in a lateral spin valve structure",
\newblock {\em Phys. Rev. B } 86, 104417 (2012)

\bibitem{Ghosh2012}
A. Ghosh, S. Auffret, U. Ebels, and W. E. Bailey,  
\newblock "Penetration depth of transverse spin current in ultrathin ferromagnets",
\newblock {\em Phys. Rev. lett.} 109, 127202 (2012)


\bibitem{Sagasta2017}
E. Sagasta, Y. Omori, M. Isasa, Y. Otani, L.E. Hueso and F. Casanova,
\newblock "Spin diffusion length of Permalloy using spin absorption in lateral spin valves",
\newblock {\em  Appl. Phys. Lett. } 111, 082207 (2017)

\bibitem{Isasa2015}
M. Isasa, E. Villamor, L. E. Hueso, M. Gradhand and F. Casanova,
\newblock "Temperature dependence of spin diffusion length and spin Hall angle in Au and Pt",
\newblock {\em Phys. Rev. B  } 91, 024402 (2015) 

\bibitem{Kimura2005}
T. Kimura, J. Hamrle, and Y. Otani,
\newblock "Estimation of spin-diffusion length from the magnitude of spin current absorption: Multiterminal ferromagnetic/nonferromagnetic hybrid structures",
\newblock {\em Phys. Rev. B  } 72, 014461 (2005)

\bibitem{Bauer2002}
G. E. W Bauer, K. M. Schep, K. Xia and P. J. Kelly,
\newblock "Scattering theory of interface resistance in
magnetic multilayers",
\newblock {\em J. Phys. D} 35, 2410 (2002)




\end{thebibliography}
\end{document}